\documentclass[reprint,aps,superscriptaddress]{revtex4-2}
\usepackage{graphicx}
\usepackage{tcolorbox}
\usepackage{bm}
\usepackage{xr}
\usepackage{float}
\usepackage{amsmath}
\usepackage{booktabs}
\usepackage[normalem]{ulem}
\usepackage{lineno}
\usepackage{siunitx}
\usepackage{hyperref}

\newcommand{\mytitle}{Identifying maximal sets of significantly interacting nodes in higher-order networks}

\begin{document}

\title{\mytitle}

\author{Lorenzo Betti}\thanks{These two authors contributed equally.}
\affiliation{Department of Network and Data Science, Central European University,  1100 Vienna, Austria}

\author{Federico Musciotto}\thanks{These two authors contributed equally.}
\affiliation{Dipartimento di Fisica e Chimica Emilio Segr\`e, Universit\`a di Palermo, Viale delle Scienze, Ed. 18, I-90128, Palermo, Italy}

\author{Federico Battiston}
\affiliation{Department of Network and Data Science, Central European University,  1100 Vienna, Austria}
\affiliation{Department of AI, Data and Decision Sciences, Luiss University of Rome, Viale Romania 32, 00197, Rome, Italy}

\author{Rosario N. Mantegna}
\affiliation{Dipartimento di Fisica e Chimica Emilio Segr\`e, Universit\`a di Palermo, Viale delle Scienze, Ed. 18, I-90128, Palermo, Italy}
\affiliation{Complexity Science Hub Vienna, Josefst\"adter Strasse 39, 1080, Vienna, Austria}

\begin{abstract}

\noindent 
Filtering methods are fundamental tools for extracting the backbone of complex networks.
Systems displaying group interactions, however, open the way to new types of filtering approaches.
Here, we introduce a statistical filtering method for higher-order networks that identifies Statistically Validated Maximal Interacting Sets -- maximal sets of nodes that consistently interact together within group interactions. 
Using properly designed benchmarks, we show that our approach is highly effective in systems where the maximal sets are likely to be diluted into interactions of larger sizes that include occasional participants.
Applications to real-world data reveal that the identified sets of nodes are characterized by higher levels of similarity and topical coherence, highlighting the ability of our method to provide new insights on the organization of real-world higher-order networks. 
\end{abstract}

\maketitle

Over the past decades, the amount and complexity of data describing natural and social systems have increased~\cite{newman2003structure}.
Networks, which are collection of nodes and edges linking pairs of them, have emerged as a powerful tool to effectively study such systems, being able to describe the interaction patterns that shape many natural and human phenomena~\cite{boccaletti2006complex}.
In many real-world systems, however, interactions can involve more than two nodes at a time. 
Higher-order networks have emerged as a natural generalization of pairwise networks to encode such many-body interactions~\cite{battiston2020networks, battiston2021physics, battiston2025higher}, where hyperedges represent interactions of any size.
This framework has been successfully applied to a variety of domains such as social~\cite{patania2017shape}, ecological~\cite{grilli2017higher}, and bio-chemical systems~\cite{jost2019hypergraph,petri2014homological}.

Filtering methods have emerged as a fundamental tool to simplify the representation of overly complex networked systems.
This task consists of pruning nodes or edges to reduce the amount of information while still maintaining the core properties of the system.
This step is required for many practical applications such as reducing connections due to noise~\cite{coscia2017network}, focus on the relevant part of the network responsible for its functioning~\cite{grady2012robust}, improving computation~\cite{mercier2022effective}, or make visualization of large networks feasible~\cite{imre2020spectrum}.
Many methods have been proposed for this task, which go under the name of backboning or network sparsification~\cite{serrano2009extracting,tumminello2011statistically,kirkley2025fast}, and have been applied to a variety of different domains~\cite{tumminello2012identification,musciotto2016patterns,musciotto2018long,li2014statistically}.
These filtering methods mainly belong to two families.
The first relies on imposing topological constraints that force to reduce the number of connections in the filtered system.
The Minimum Spanning Tree belongs to this family, where a system of $n$ connected nodes is reduced to a tree with $n-1$ links~\cite{mantegna1999hierarchical}, along with extensions involving new topological constraints~\cite{tumminello2005tool,massara2016network} and statistical approaches~\cite{tumminello2007spanning,musciotto2018bootstrap}.
The second family is based on dropping connections that are compatible with an appropriate null model~\cite{serrano2009extracting,tumminello2011statistically,coscia2017network}.
These models generate ensembles of randomized replicas of the system while preserving some key features of the original network.
An example is the disparity filter, which prunes edges based on their weights~\cite{serrano2009extracting}.
This statistical filter, along with similar methods~\cite{tumminello2011statistically,coscia2017network}, naturally account for the heterogeneity and multiscale connectivity pattern observed in real-world networks~\cite{barrat2004architecture}.

In higher-order networks, developing powerful methods for filtering represents an even more pressing need, given the richness of information encoded in multi-body interactions~\cite{musciotto2021detecting, landry2024filtering}.
The multi-body nature of higher-order interactions not only requires to adapt existing methods developed for pairwise networks but also opens the door to entirely new methodological approaches that are not direct generalizations of existing tools.
A stream of methods -- falling under the name of reducibility -- have approached the issue of dimensionality reduction of higher-order systems by separating interactions according to their order (number of interacting units per interaction), assessing which interaction orders are more relevant at a global level, and retaining or pruning entire interaction orders based on their structural~\cite{kirkley2025structural} or dynamical~\cite{lucas2025reducibility} relevance.
Other approaches take a local perspective, more similar to traditional filtering approaches, such as the Statistically Validated Hypergraphs (SVH), which, for each interaction size, validates statistically each single hyperedge based on a local null model~\cite{musciotto2021detecting}.
These approaches act by pruning existing hyperedges, with the resulting higher-order network composed exclusively of a subset of the interactions that were present originally.
However, when interactions involve multiple nodes, the internal composition of a hyperedge becomes itself an object of interest. 
This consideration opens the door to the possibility of identifying significantly interacting sets of nodes \textit{within} observed hyperedges, possibly across orders of interactions, in contrast to validating entire hyperedges only.
Filtering approaches taking this new and broader perspective are currently still lacking.

In this work, we propose a filtering methodology for higher-order complex systems capable of extracting Statistically Validated Maximal Interacting Sets (SVMIS) from  collectively interacting groups of nodes across interactions of different sizes.
Our method relies on a null model designed for repeated higher-order interactions, that is able to identify the maximal sets of nodes that consistently interact collectively, and discard the nodes that co-interact with them only occasionally, while inheriting the capability of preserving network heterogeneity typical of other network filtering approaches~\cite{tumminello2011statistically,musciotto2021detecting}.
Given the combinatorial nature of the problem at hand, we propose an approximation method to calculate the statistical significance of sets of nodes that overcomes this computational bottleneck (see Section~\ref{app:approx_pval} of the Appendix) .
We assess the method’s performance on a class of synthetic benchmarks and demonstrate its ability to extract meaningful structure in empirical data, highlighting its general applicability to higher-order complex systems.

\section{Filtering method}

Given a collection of group interactions, we aim at developing a method that identifies sets of nodes that consistently interact with one another across collective interactions.
In other words, we want to identify those sets of nodes that repeatedly appear together \textit{within} group interactions more often than expected by chance under an appropriate null model.
Unlike traditional filtering methods that prune nodes or entire interactions, our method focuses on filtering nodes \textit{within} group interactions to reveal statistically meaningful sets of interacting nodes.
For example, in a retail setting where group interactions correspond to shopping baskets, consider a set of $m$ products that are functionally related, such as yeast, flour, mozzarella and tomato sauce, which are needed to make pizza.
Although these products may often be bought along with other, potentially unrelated products because of the heterogeneity of shopping habits, classic filtering methods would merely prune \textit{entire} shopping baskets, thus overlooking the recurring co-occurrence of sets of items \textit{within baskets}.
Instead, our method aims at detecting such sets of $m$ products that systematically co-appear in baskets of size $m$ or larger, out of all the ones that are only occasionally co-purchased with them.

\begin{figure}[ht!]
	\centering
	\includegraphics[width=0.3\textwidth]{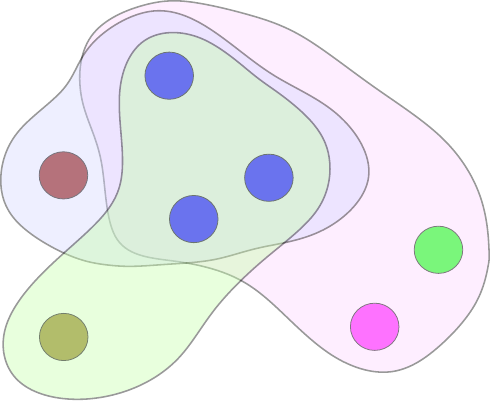}
	\caption{{\bf Statistically validated sets}.
    Hyperedges (shaded areas) that contain a statistically validated set of nodes (blue nodes).
    The three validated nodes consistently appear together in different hyperedges, which also contain other nodes.
    \label{fig:sketch}}
\end{figure}

In the following we adopt the formalism of higher-order networks~\cite{battiston2020networks} to develop our method.
Within this formalism, we represent systems as hypergraphs, where higher-order interactions between an arbitrary number of nodes are encoded as hyperedges.
A hyperedge of size $n$ corresponds to an interaction among $n$ nodes (e.g., $n$ products in the same basket). 
Our purpose can be framed as transforming a collection of hyperedges into a set of statistically validated sets, where each of them represents a group of consistently interacting nodes (Fig.~\ref{fig:sketch}).

It is worth stressing that the validated node sets are not necessarily a subset of the hyperedges of the original hypergraph because $m$ nodes consistently appearing together may always occur within hyperedges of size $m$ or larger.
This aspect distinguishes our method from more traditional filters developed for higher-order networks like SVH~\cite{musciotto2021detecting} and other related approaches~\cite{ceria2025relevance,landry2024filtering}.
Additionally, our method is also related to the notion of nestedness in hypergraphs, which quantifies the extent to which hyperedges are entirely embedded within larger ones~\cite{lotito2022higher,larock2023encapsulation,landry2024simpliciality,lamata2025hyperedge}.
In nested systems, subsets of nodes that repeatedly appear across multiple interaction sizes are natural candidates to emerge as validated interacting sets of nodes. 
From this perspective, our method can be interpreted as identifying the local node-level cores that underpin broader nestedness patterns. 
Conversely, significantly interacting nodes can also arise in the absence of strict containment relations between hyperedges, as illustrated in Fig.~\ref{fig:sketch}, showing that our method and measures of nestedness capture complementary aspects of the organization of higher-order networks.

In what follows, we will call the filtered sets of interacting nodes Statistically Validated Maximal Interacting Sets (SVMIS).

More concretely, our method works as follows.
We consider a hypergraph with $N$ hyperedges, and given a set of $m$ nodes $S=\left\{i_1, i_2, \dots, i_m \right\}$, we define $N_{i_1i_2\dots i_m}$ as the number of times these nodes co-appear in hyperedges of size $m$ or larger.
In case of a weighted hypergraph, where edge weights represent the multiplicity of group interactions, $N_{i_1i_2\dots i_m}$ is defined by accounting for these multiplicities.
Let also $N_i$ be the number of hyperedges node $i$ participates in, regardless of the size.
Our null hypothesis states that each node $i$ independently selects the hyperedges it participates uniformly at random, while keeping $N_i$ fixed.
Equivalently, in the bipartite representation of a hypergraph -- where one layer refers to nodes and the other to hyperedges -- this null hypothesis corresponds to a configuration-model randomization that fixes the degree in the node layer (i.e. $N_i$) and rewires their connections uniformly at random~\cite{squartini2015breaking}.
The task of identifying statistically validated interacting sets of nodes corresponds to computing the probability $p(N_{i_1 i_2 \dots i_m})$ that a given set $S$ of $m$ nodes co-occurs in a sufficient number of hyperedges, as compared to the expectation according to the null model.

To compute the probability that nodes in $S$ co-occur inside $N_{i_1i_2\dots i_m}$ hyperedges, we apply a hierarchical procedure.
For the sake of illustration, we first derive the case where the set $S$ is composed of three nodes ($m=3$) before providing the general expression.
The probability of observing the first two nodes 
co-occurring in exactly $X_{12}$ hyperedges follows a hypergeometric distribution:
\begin{align}
H(X_{12}|N,N_{1},N_{2}) = \frac{ \binom{N_{1}}{X_{12}} \binom{N-N_{1}}{N_{2} - X_{12}} }{ \binom{N}{N_{2}} }.
\end{align}
Conditioned on $X_{12}$, the probability of observing the three nodes occurring in $N_{123}$ hyperedges is again described by a hypergeometric distribution:
\begin{align}
    H(N_{123}|N,X_{12},N_{3}) = \frac{ \binom{X_{12}}{N_{123}} \binom{N-X_{12}}{N_{3} - N_{123}} }{ \binom{N}{N_{3}} }.
\end{align}
Then, by summing over all feasible values of $X_{12}$, we obtain the probability of observing $N_{123}$ hyperedges where the three considered nodes 
are all present at the same time:
{\footnotesize
\begin{align*}
		p(N_{123}) &= \sum_{X_{12}} H(X_{12}|N,N_{1},N_{2})\times  H(N_{123}|N,X_{12},N_{3}) \\
						 &= \frac{1}{\binom{N}{N_{2}}\binom{N}{N_{3}}}\sum_{X_{12}} \binom{N_{1}}{X_{12}}\binom{N-N_{1}}{N_{2}-X_{12}}\binom{X_{12}}{N_{123}}\binom{N-X_{12}}{N_{3} - N_{123}},
					 \stepcounter{equation}\tag{\theequation}\label{eq:prob3}
    \end{align*}
}
where $X_{12} \in \left[N_{123}, \min(N_{1}, N_{2}) \right]$.
To obtain the probability for an arbitrary number of nodes $m$, we iterate further the procedure and we get: 
{\footnotesize
	\begin{align*}
		p(N_{12\dots m}) = &\sum_{X_{12}} H(X_{12}|N,N_{1},N_{2})\times \\
							 &\times \sum_{X_{123}} H(X_{123}|N,X_{12},N_{3})\times... \nonumber \\
							 &...\times\sum_{X_{1...m-1}}H(X_{1...m-1}|N,X_{1...m-2},N_{m-1})\times\\
							 &\times H(N_{1\dots m}|N,X_{1...m-1},N_{m}).
\stepcounter{equation}\tag{\theequation}\label{eq:probn}
\end{align*}
}
where $X_{1 2 \dots r}$ denotes the number of hyperedges containing the first $r$ nodes, and the summation limits are given by:
\begin{align}
    X_{1 2 \dots r} &\in \left[ N_{1 \dots r+1}, min(X_{1 \dots r-1}, N_{r}) \right] .
\end{align}

Note that Eq.~\ref{eq:probn} is symmetrical for permutations of the order of the nodes, making this expression suitable for any set of nodes regardless of the order and thus $p(N_{12\dots m}) = p(N_{i_1 i_2\dots i_m})$ for any permutation of the nodes in $S$.
Although Eq.~\ref{eq:probn} may appear similar to Eq.~3 in the reference introducing the SVH method~\cite{musciotto2021detecting}, there is a crucial difference.
For SVH, since hyperedges are validated for each order separately, the null expectation is built by considering all hyperedges of size $m$. 
Instead, in our approach we look at all hyperedges of size $m$ or larger, thus making it possible to filter nodes \textit{within} hyperedges.
In summary, Eq.~\ref{eq:probn} is an iterative composition of hypergeometric intersections~\cite{wang2015efficient}.

Having derived an expression for $p(N_{i_1i_2\dots i_m})$, we can finally assess the statistical significance of a set of $m$ nodes against a threshold of statistical significance $\alpha$.
The corresponding p-value of the given set of nodes can be obtained through the survival function:
\begin{equation}
	\label{eq:pvalue_3}
		p(X\geq N_{i_1i_2\dots i_m}) = 1 - \sum_{x=0}^{N_{i_1i_2\dots i_m} - 1} p(x).
\end{equation}
For the rest of this paper, we set $\alpha=0.01$.
We apply a Benjamini-Hochberg correction for false discovery rate to account for the large number of tested hypotheses~\cite{benjamini1995controlling} (see Section~\ref{app:bh_correction} of the Appendix for details).

Since we are interested in finding statistically validated maximal interacting sets, we are in principle interested in assessing all possible combinations of nodes of any size.
This approach would be computationally unfeasible because the number of such sets grows as the Bell number of the number of nodes (i.e. faster than exponential).
However, since sets of nodes that never appear together in any hyperedge cannot represent meaningful sets of interacting nodes ($N_{i_1i_2\dots i_m} = 0$ making $p(X\geq N_{i_1i_2\dots i_m}) = 1$), we restrict the set of nodes to test to those that appear at least once together within a hyperedge (i.e. such that $N_{i_1i_2\dots i_m} \geq 1$), which reduces the search space.
Moreover, we focus on maximal interacting sets.
Namely, once a set of $m$ nodes is found to be statistically validated, we do not test any of the subsets of such $m$ nodes.
Note that those subsets of a statistically validated set should not necessarily be considered as significantly interacting as well.
Thus, our approach mainly focuses on the maximally large set of interacting nodes and can thus neglect other smaller significant sets within larger ones.

To further ease this computational bottleneck, we propose an approximation for the computation of the p-value by approximating Eq.~\ref{eq:probn} with a binomial distribution, which tends to overestimate the exact p-value (see Section~\ref{app:approx_pval} of the Appendix for details).
This approximation recalls the one proposed in Ref.~\cite{kobayashi2019structured}.
In all the applications that we are going to showcase in the following, we will use the approximated p-value.

Given those considerations, for each size $m$, we proceed by testing sets of nodes from largest to smallest, where the largest $m$ corresponds to the size of the largest observed hyperedge.

\section{Statistically validated maximal interacting sets}

\subsection{Definition and investigation of benchmarks}

To test the effectiveness of our approach in identifying sets of nodes that interact frequently within hyperedges, we design benchmark hypergraphs with implanted sets of nodes representing the interacting groups that the method should recover.
Starting from these implanted sets, we generate hyperedges by eventually ``diluting'' them with additional nodes selected at random.
This controlled setting is not intended to reproduce all structural features of empirical hypergraphs, but rather to showcase the context in which SVMIS is expected to be useful: identifying recurrent core groups that may appear within larger higher-order interactions.

We create benchmark hypergraphs made of $N$ nodes in which we implant $T$ sets for each size $n$.
The number of implanted sets is controlled through a density parameter $d$ such that, for each realization of the benchmark, we set $T=d \binom{N}{2}$ to make the number of implanted sets grow with $N$ while keeping benchmarks of different sizes comparable.
In this work, we explored $N=\{100,200,500\}$ and $d \in \{0.005, 0.01, 0.02\}$, with $n$ ranging in the interval $[2,4]$.
For each implanted set, we create $l$ hyperedges of size up to $n_{max}$, where $l$ is sampled from a Binomial distribution with $p_{bin}=0.5$ and $n_{bin}=20$.
Each hyperedge contains the $n$ nodes of the implanted set plus $n'$ additional nodes, with $n'$ sampled from the discrete uniform distribution $U(0, n_{max} - n)$.
This sampling yields hyperedges that either coincide with the implanted sets (when $n'=0$) or are diluted with randomly chosen nodes (when $n'>0$).
In our simulations, we set $n_{max}=6$ to obtain a moderate dilution of the implanted sets into larger hyperedges, thus having the maximum hyperedge size in $[8, 10]$ depending on the value of $n$.
In this way the $n$ nodes of a set interact $l$ times in different hyperedges, which always contain the $n$ core nodes plus a set of randomly selected ones.
Moreover, we introduce a parameter $c$ that controls the level of closure at each group size.
Specifically, for an implanted set of size $n$, $c$ represents the closure probability, defined as the probability that all lower-order interactions among the $n$ nodes are fully nested within the group~\cite{lotito2022higher,kirkley2025structural,landry2024simpliciality}.
For example, when $n=3$, the closure probability $c$ determines the probability of including all the three possible pairwise interactions among the nodes.

We compare the results of SVMIS against the Statistically Validated Hypergraphs (SVH) method~\cite{musciotto2021detecting}.
SVH was developed for hypergraphs but it only prunes entire hyperedges, in contrast to SVMIS.

To validate our results, we compute the True Positive Rate (TPR), that quantifies the ratio of implanted sets that we are able to identify, and the False Discovery Rate (FDR) that quantifies the ratio of erroneously detected implanted sets that the different methods select.
Specifically, TPR=$\frac{TP}{TP+FN}$ and $FDR = \frac{FP}{FP+TP}$, where TP are the True Positives (i.e. significant sets corresponding to the originally implanted sets), FN are the False Negatives (i.e. implanted sets that are not identified) and FP are the False Positives (sets detected as significant by the method that are not among the originally implanted sets).
We provide sensitivity analyses for different values of $\alpha$ in Section~\ref{app:sensitivity_alpha} of the Appendix.

Fig.~\ref{fig:bench}a shows TPR as a function of $c$ for N=200 and $d=0.005$, but the observed behavior is independent of the specific values of N and $d$.
Each point is the median of 100 benchmark realizations.
The color bands represent the interval between the $10$-th and the $90$-th percentile.
The performance of SVMIS and SVH is not affected by $c$, as both are methods designed to take into account the higher-order nature of interactions.
In fact, if 3 nodes $i,j,k$ constitute a clique of (significant) pairwise interactions but never interact altogether, SVMIS and SVH will not statistically validate the corresponding set because $N_{ijk}=0$ even if $N_{ij}$, $N_{ik}$ and $N_{jk}$ are not.
As a consequence, SVMIS and SVH will not miss $\{ij\}$, $\{jk\}$ and $\{ik\}$ as significant set.
We also observe that SVMIS significantly outperforms SVH.
This happens because in our benchmark hypergraphs we allow the $n$ nodes of an implanted set to eventually interact within larger hyperedges.
While SVMIS counts all the hyperedges in which the $n$ nodes appear, the SVH approach misses all those in which there are additional, randomly selected nodes.
For this reason, SVH is more likely to underestimate the significance of a set and thus have a lower value of TPR.
Fig.~\ref{fig:bench}b plots the FDR as a function of $c$.
SVH and SVMIS almost never detect wrong sets as they distinguish between interactions of different orders.

To better characterize the difference in performance between SVMIS and SVH, we modify the benchmark by enforcing the control on the extent to which the implanted sets are diluted within larger hyperedges.
We introduce a parameter $f$ that regulates the fraction of times in which the $n$ nodes of an implanted set appear together with additional nodes (i.e. in hyperedges of size larger than $n$).
Specifically, for each implanted set of size $n$, we create $l$ hyperedges such that (i) a fraction $f$ of newly created hyperedges are generated by diluting the implanted set with other $n'$ random nodes, with $n'$ sampled from $U (1, n_{max} - n)$, and (ii) the remaining fraction $1-f$ consists of hyperedges equal to the implanted set itself (i.e. $n'=0$).
The change on the lower bound of $n'$ from 0 to 1 ensures that $f$ fully controls the fraction of implanted sets diluted within larger hyperedges.
The larger $f$, the more we expect SVH to underestimate the number $N_{i_1i_2\dots i_m}$ of collective interactions and thus will likely fail to validate the true sets.
In Fig.~\ref{fig:bench}c we show TPR as a function of $f$.
Indeed, we see that SVH significantly worsens its performance when $f$ increases.
Fig.~\ref{fig:bench}d shows the behavior of FDR as a function of $f$.
The FDR for SVH remains low not because SVH is accurate, but because the method detects almost no positives (see the TPR in Fig.~\ref{fig:bench}c).

It is worth noting that the TPR values shown in Fig.~\ref{fig:bench} represent a lower bound on the performance achievable with the exact p-value computation. 
Since the approximate p-values tend to overestimate the exact ones (see Section~\ref{app:approx_pval} of the Appendix), the approximation exclusively introduces additional false negatives (i.e. implanted sets that fail to be validated) without affecting the false discovery rate. 
The consistently high TPR observed across all benchmark settings therefore confirms that the approximate p-value computation offers a favourable trade-off: it makes SVMIS computationally feasible while preserving strong performance in identifying significantly interacting sets of nodes within higher-order interactions.

\begin{figure*}[ht!]
	\centering
	\includegraphics[width=0.95\textwidth]{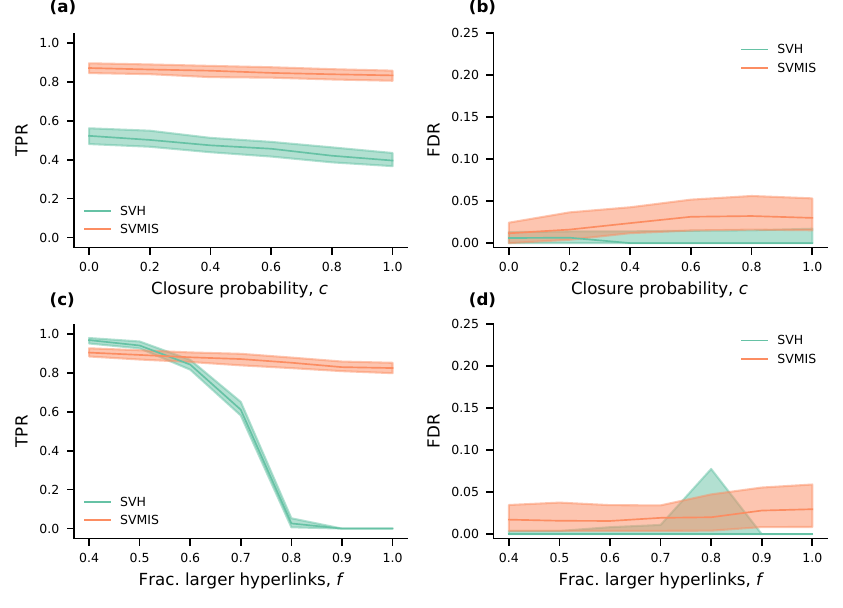}
	\caption{{\bf Benchmark analysis} (a) True Positive Rate (TPR, y-axis) between the implanted sets and those detected by Statistically Validated Hypergraphs (SVH, sea-green line) and
    Statistically Validated Maximal Interacting Sets (SVMIS, orange line) as a function of the closure probability $c$ with N=200 and $d=0.005$.
    Solid lines refer to the median across 100 realizations of the benchmark and transparent bands to the interval between the $10$-th and the $90$-th percentile. 
    (b) False Discovery Rate (FDR, y-axis) for SVH and SVMIS as a function of $c$ for the same benchmark realizations. 
    (c) TPR (y-axis) for SVH and SVMIS as a function of the fraction of larger hyperedges $f$ over 100 benchmark realizations. 
    (d) FDR (y-axis) for SVH and SVMIS as a function of the fraction of larger hyperedges $f$ over the same benchmark realizations of (c).\label{fig:bench}}
\end{figure*}

In sum, the benchmark analysis reveals that SVMIS is able to identify sets of significantly interacting nodes within larger group interactions.
Moreover, the comparison against SVH confirms that these filtering approaches differ from SVMIS as a result of the different problem they aim to address.
In the following, we apply SVMIS to real-world data.

\subsection{Application to real data: Baskets of co-purchased goods at Walmart}

\begin{figure*}[ht!]
    \centering
    \includegraphics[width=0.99\textwidth]{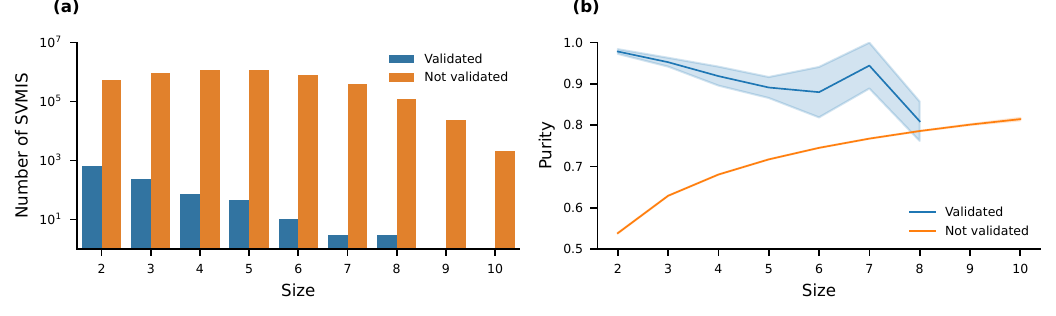}
    \caption{\label{walmart}{\bf Significant co-purchases at Walmart.} 
    (a) Counts of validated sets (blue bars) and not validated ones (orange bars) as a function of size.
    The sum of blue and orange bars constitutes the total number of tested sets. 
    (b) Solid lines represent the average purity and transparent bands are the standard error for validated (blue) and not validated (orange) sets.}
\end{figure*}

We analyse a dataset that tracks 69,906 co-purchases of 88,860 products at Walmart~\cite{amburg2020clustering}, available as part of the data repository Hypergraphx-data~\cite{lotito2026hypergraphxdata}.
This system can be mapped into a hypergraph where hyperedges represent co-purchases.
In this context, we use the SVMIS approach to extract sets of products that are inherently bought together. 
When extracting the Statistically Validated Maximal Interacting Sets we exclude from the hypergraph all baskets of size larger than 10 or equal to 1.
Indeed, each large basket of size $k$ which is not statistically significant produces a large number of candidate sets of smaller size $k'$ - specifically it produces $\binom{k}{k'}$ candidates.
Given that larger baskets have a small number of co-occurrence, we cut the system at size 10 in order to avoid the explosion in the number of hyperedges with low chance of getting validated.

Fig.~\ref{walmart}a shows the number of validated (blue) and not validated (orange) sets for each size.
The number of validated sets exponentially decays with size, and, for large $n$, it is several orders of magnitude smaller than the sets that are observed at least once but are not rejecting the null hypothesis.

The dataset also reports the category of each product.
The categories are: Clothing, Shoes \& Accessories; Electronics and Office; Home, Furniture \& Appliances; Home Improvement \& Patio; Baby; Toys, Games, and Video Games; Food, Household \& Pets; Pharmacy, Health \& Beauty; Sports, Fitness \& Outdoors; Auto, Tires \& Industrial; Other. 
By using this information, we define a proxy of ``purity" of co-purchased goods.
Purity is a measure that goes from 0 (if all products in a group belong to different categories) to 1 (if they all belong to the same category), and is defined as $P=\frac{n}{n-1} (1- \frac{C}{n})$, where $n$ is the size of the set and $C$ is the number of unique categories to which the $n$ products belong.
Fig.~\ref{walmart}b shows the ``purity'' for the validated (blue) and not validated (orange) sets as a function of their size.
In the figure, the solid lines plot the mean and the transparent bands the error of the mean.  
These results show that, for each size, the validated sets identified by SVMIS correspond to sets of items that are more homogeneous with respect to category than the ones that are not validated.

\subsection{Co-traded stocks belonging to the S\&P100 index}

We investigate a dataset tracking tick-by-tick transactions performed by 96 stocks that belong to the S\&P100 index and are traded on any US venue in the month 12/2020~\cite{firstrate}.
The data comes with a temporal resolution of 1 ms.
By applying our method to this dataset, we detect SVMISs of stocks that are co-traded in the investigated period in excess to what expected by a null model taking into account the heterogeneity observed in stock trading.
Again, before extracting the SVMISs we exclude from the hypergraph all hyperedges larger than 10 stocks or equal to 1.
We perform our analysis at a daily scale and then aggregate the results obtained for different days.

\begin{figure*}[ht!]
    \centering
    \includegraphics[width=0.95\textwidth]{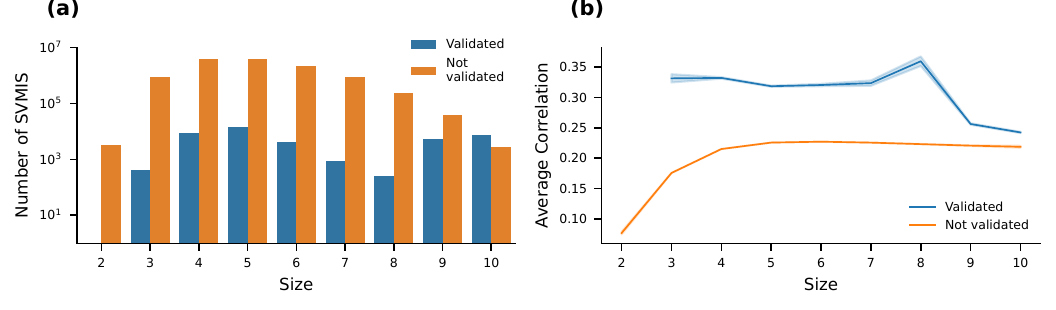}
    \caption{\label{fig:sp}{\bf Co-traded stocks from S\&P100 index.} 
    (a) Counts of validated sets (blue bars) and not validated ones (orange bars) as a function of size.
    The sum of blue and orange bars constitutes the total number of tested sets. 
    (b) Solid lines represent the average correlation and transparent bands are the standard error among stocks in the validated (blue) and not validated (orange) sets. 
    }
\end{figure*}

Fig.~\ref{fig:sp}a shows the number of validated (blue) and not validated (orange) sets for each size.
Fig.~\ref{fig:sp}b shows the mean correlation (solid line) and its error (transparent bands) between couples of stocks in validated (blue) and not validated (orange) sets.
Correlation is computed on time series of financial returns extracted at a resolution of 1 minute.

By identifying sets of stocks that are co-traded significantly more often than expected by chance, our method reveals higher-order structures in price formation characterized by significantly higher average price correlations compared to non-validated sets. 
This provides empirical confirmation to the theory that high-frequency collective activity contributes significantly to price synchronization across related assets~\cite{gerig2015high}, linking the algorithmic driven co-occurrence of market interactions directly to the emergence of coordinated price dynamics.

\subsection{Food recipes}
Finally, we apply our method on a dataset representing food recipes from different cuisines around the world~\cite{culinarydb}.
We used the preprocessed dataset provided in Ref.~\cite{caprioli2025networks}, which can be represented as a hypergraph where ingredients correspond to nodes and recipes (i.e. groups of ingredients) to hyperedges.
After excluding recipes composed of one single ingredient or more than 10 ingredients, the resulting hypergraph has 580 nodes and \num{32291} hyperedges.

\begin{figure*}
    \centering
    \includegraphics[width=0.75\linewidth]{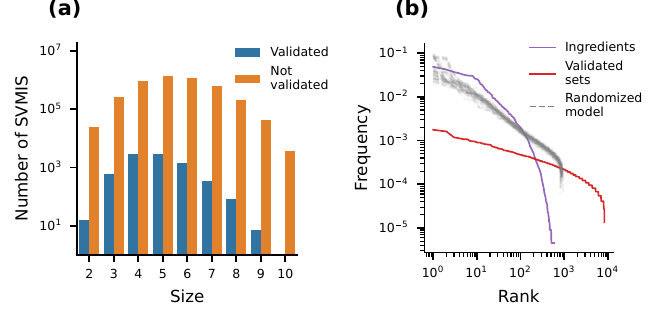}
    \caption{
    \textbf{Food recipes.} (a) Counts of validated sets (blue bars) and not validated ones (orange bars) as a function of size.
    The sum of blue and orange bars constitutes the total number of tested sets.
    (b) Frequency-rank plots of ingredients across recipes (violet line), validated sets across real recipes (red line) and across randomized recipes (grey lines).
    The randomized ones represent the result from 50 independent realizations of the randomized model.
    }
    \label{fig:recipes}
\end{figure*}

Fig.~\ref{fig:recipes}a shows the number of validated (blue) and not validated (orange) sets as a function of their size.
Validated sets are concentrated at intermediate sizes and display a rapid decay for larger sizes.
This suggests that world cuisines are structured around relatively small and recurrent combinations of ingredients.

To illustrate the non-trivial nature of the validated sets identified by SVMIS, we examine their frequency across recipes.
Fig.~\ref{fig:recipes}b shows the frequency-rank plot of validated sets (red line), which spans almost three orders of magnitude and exhibits finite-range scaling behavior.
In contrast, the frequency-rank plot of ingredients is distributed more narrowly (violet line).

We compare this observation against a simple randomized model where, for each cuisine, we assign ingredients to recipes randomly while preserving ingredient usage frequency and the distribution of the number of ingredients per recipe (i.e. a bipartite configuration model in the ingredient-recipe bipartite network).
Although displaying a finite-range scaling behavior, the slope is steeper compared to the one observed in the real system (grey lines).

In sum, these observations illustrate that SVMIS identifies non-trivial recurrent structures in the recipe hypergraph, which are not captured by mere ingredient frequency statistics and are not compatible with random combination of ingredients.
In this sense, the method reveals non-trivial organization at the level of ingredient combinations. 

\section{Discussion}

The recent surge in interest towards higher-order networks, together with the increased availability of large amount of data on several real world systems, have led to the introduction of the first filtering tools for hypergraphs~\cite{kobayashi2019structured,musciotto2021detecting}.
However, the task of detecting statistically validated units of collectively interacting elements is multi-faceted, and the multidimensional nature of higher-order interactions calls for the design of different validation approaches.
The existing methods have focused on the extension of validating approaches from pairwise interactions to hyperedges~\cite{kobayashi2019structured} and the detection of statistically validated group interactions that occur only at a specific size~\cite{musciotto2021detecting}.
Here, we introduced a different and more general framework for the detection of Statistically Validated Maximal Interacting Sets, which represents sets of nodes that appear together within group interactions. 
The computational efforts needed for the statistical validation are demanding but accessible with ordinary PCs for groups up to a size of about 10 elements.
To speed up the computation we have proposed an approximated procedure that works well in a large number of cases as shown in our investigation of benchmarks.
Using properly designed hypergraph benchmarks, we show that our approach is highly effective on higher-order systems (i) in which informative interactions among $n$ core nodes are likely to be diluted into larger hyperedges that feature occasional nodes and (ii) with a high degree of closure.
We also show how, on real data, our approach allows to detect significantly interacting sets whose nodes are characterized by high levels of similarity, which might help in shedding light on the generative process of the underlying hypergraphs.

\subsection*{Code Availability}
Data and scripts needed to replicate the results are available at the GitHub repository \url{https://github.com/musci8/SVH}.
The code to extract Statistically Validated Maximal Interacting Sets is available as part of the library Hypergraphx~\cite{lotito2023hypergraphx}.

\subsection*{Acknowledgments}
F.B.~acknowledges support from the Austrian Science Fund~(FWF) through projects~10.55776/PAT1052824 and 10.55776/PAT1652425.

\pagebreak
\clearpage
\widetext
\appendix
\begin{center}
\textbf{\large Appendix}
\end{center}
\setcounter{equation}{0}
\renewcommand{\theequation}{A\arabic{equation}}

\setcounter{figure}{0}
\renewcommand{\thefigure}{A\arabic{figure}}

\setcounter{table}{0}
\renewcommand{\thetable}{A\arabic{table}}

\setcounter{section}{0}
\renewcommand{\thesection}{A\arabic{section}}

\section{Approximated p-values}\label{app:approx_pval}

Extracting the probability in Eq.~4 of the main text requires the convolution of $m-1$ instances of hypergeometric distributions, each of which is built through the combination of several binomial coefficients.
For this reason, this task becomes computationally intense for larger $m$.
On top of this, our approach may require the computation of several p-values over the same nodes, as any set of $m$ nodes which does not represent a significant interacting set is split into $m$ candidate sets of size $m-1$.
If not significant, each of these sets is further split into candidate sets of size $m-2$, and so on, quickly causing the number of computationally intense p-value that need to be computed to explode.
To overcome this computational bottleneck, we have tested an approximated version of the p-value.
Eq.~4 of the main text represents the probability that $m$ nodes appear in $N_{12...m}$ hyperedges after each of them has randomly sampled its hyperedges from an urn.
As the random sample performed by each node is independent from the others, we can approximate this estimation using the binomial distribution, turning Eq.~4 of the main text into:
\begin{equation}
	\label{eq:approx}
	p(N_{12...m}) = \binom{N}{N_{12...m}}p^{N_{12...m}}(1-p)^{N-N_{12...m}},
	\end{equation}
	where $p=\prod_{i=1}^{m+1}\frac{N_i}{N}$.
In other words, we are estimating the probability of observing $N_{12...m}$ hyperedges (out of the $N$ total hyperedges observed) that involve the $m$ nodes, where each hyperedge has a probability of being observed which is the product of the normalized number of hyperedges $\frac{N_i}{N}$ in which each node appears.
Our approximated p-value recalls the one proposed in Ref.~\cite{kobayashi2019structured}, as both approaches use the binomial distribution.
However, here we model differently the probability of observing a collective interaction.
In Ref.~\cite{kobayashi2019structured}, given the specific nature of the investigated data that only tracks temporal pairwise interactions, the $p$ in the binomial distribution is obtained by multiplying the probabilities of observing all pairwise links that constitute a clique of the same size of the tested hyperedge.
Here, we obtain it by multiplying the normalized number of hyperedges of each node involved in the interaction.
The approximation of Eq.~\ref{eq:approx} lies in the fact that we do not take into account anymore the $m$-dimensional array of individual activities of each node in the hyperedge $\{N_i\}$, but we consider the normalized 1-D product between them. 
	
\begin{figure*}[ht!]
	\centering
	\includegraphics[width=0.95\textwidth]{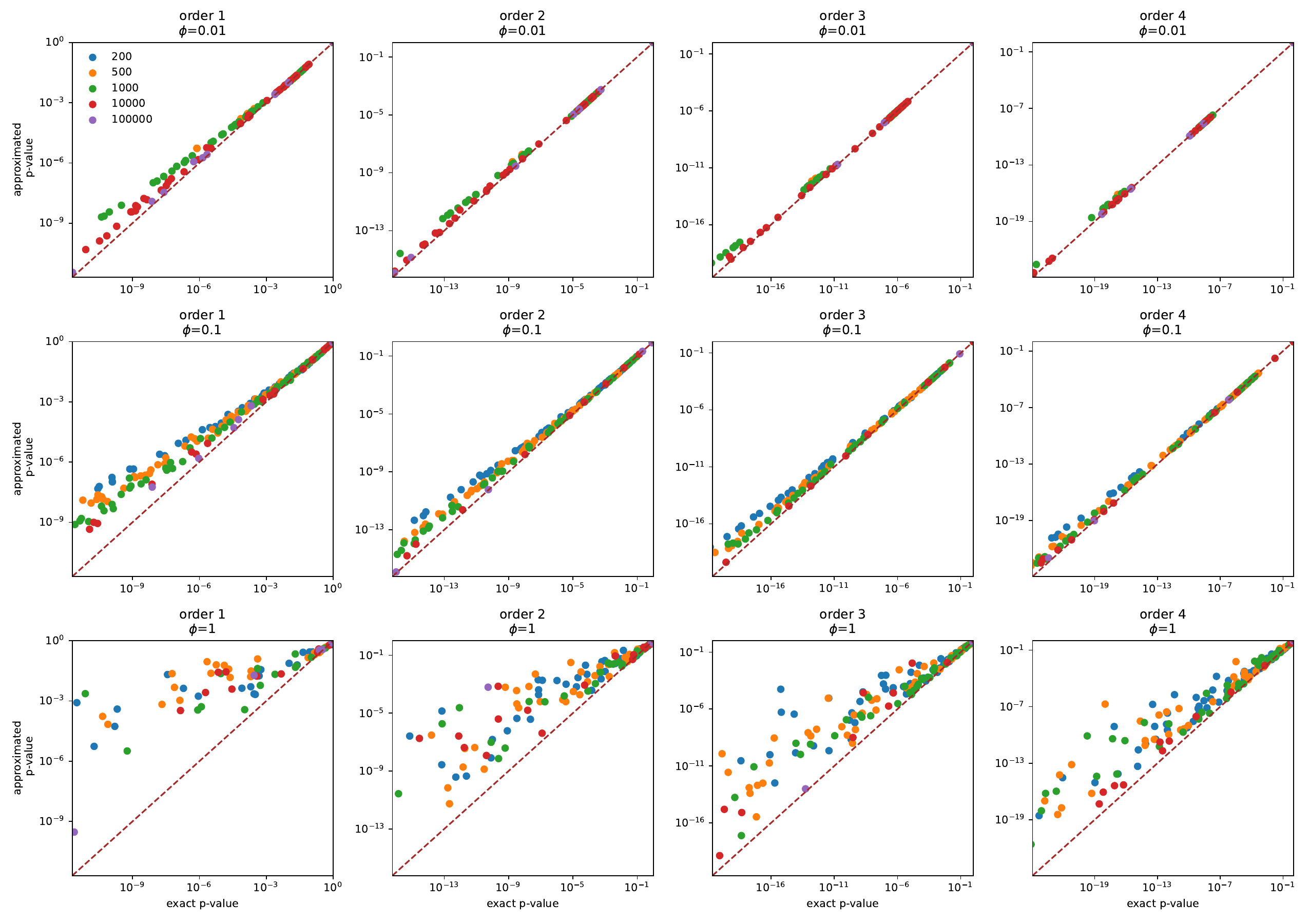}
	\caption{{\bf Comparison between exact and approximated p-values}.
    Each panel is a scatter plot between the exact p-values extracted from Eq.~4 of the main text (x-axis) and the ones extracted from Eq.~\ref{eq:approx} (y-axis) in our simulations.
    The colors of the dots reflect the number of hyperedges $N$ used in the simulation.
    Panels in the same columns display results for sets of the same size (2, 3, 4, and 5 from left to right).
    Panels in the same row display results with different values of $\phi$ (on top $\phi=0.01$, in the center $\phi=0.1$ and on the bottom $\phi=1$.) \label{fig:pvalue}}
\end{figure*}

We thus explored the precision of this approximation in different scenarios (Fig.~\ref{fig:pvalue}).
To do so, we randomly generated groups of nodes of different order with different values of $\{N_i\}$ (out of $N$ total hyperedges) and different values for the number of collective hyperedges $N_{1 2\dots m}$.
Specifically, for each size $n$ we extracted $n$ values of $N_i$ from the uniform distribution $U(1,\phi N)$, where $\phi$ is the fraction of maximum number of hyperedges out of the total $N$ that we allow a node to appear in.
We then extracted $N_{1 2\dots m}$  from the uniform distribution $U(1, min({N_i}))$.
We then used the values $(N_{1,2,...,m},{N_i},N)$ to extract a p-value with the exact approach used in Eq.~4 of the main text and the approximated one of Eq.~\ref{eq:approx}.
We varied the size of the sets to validate in the interval $[2, 5]$.
Each column of Fig.~\ref{fig:pvalue} presents results with a different size.
We also varied the total number of interactions $N$ using the values $N=[200,500,1000,10000,100000]$.
For each panel of the Figure we show results for different $N$ with dots of different colors.
Moreover, we varied the fraction $\phi$ of maximal activity allowed to each node using the values $\phi=[0.01,0.1,1]$.
Each row of Fig.~\ref{fig:pvalue} presents results with a different value of $\phi$.
We find that across the different settings, the approximated p-value tends to overestimate the effective probability that the number of interactions within a group is compatible with our null hypothesis.  
In particular, this overestimation occurs precisely in the regime relevant for validation, namely when the observed number of collective interactions exceeds the expectation under the null model.
This can lead to False Negatives, i.e. to miss groups that are significant but for which we overestimate the actual p-value, resulting in a reduction of the True Positive Rate without inflating the False Discovery Rate.
However, this overestimation is pronounced only for values of $\phi$ close to 1, which means cases in which a system is made of a small amount of nodes each of which is responsible, on average, of a significant portion of the complete number of hyperedges.
In cases in which a single node is at maximum present in a smaller fraction of the overall interactions, the approximation works well and the p-values computed using the two approaches fall on the diagonal.

In all the applications that we showed in the main text, we used the approximated p-value. 
We verified that in all empirical networks $\phi > 0.1$ for less than 5\% of nodes.

\section{Correction for multiple hypothesis tests}\label{app:bh_correction}

The SVMIS method requires assessing the statistical significance of all possible sets of nodes of any size.
For example, given $N$ nodes and size $m$, the number of test is equal to all possible sets of size $m$ given $N$ nodes, which is equal to $\binom{N}{m}$.
When many hypotheses are tested simultaneously, the probability of obtaining false positives increases substantially (i.e. validating a set just because of random chance).
Without correcting for this effect, a small significance level $\alpha$ can result in a high number of spuriously validated sets.

To address this issue, we apply the Benjamini-Hochberg (BH) procedure~\cite{benjamini1995controlling}.
This procedure controls the false discovery rate -- the expected proportion of falsely rejected null hypotheses among all rejected ones.
The BH procedure works by sorting all computed p-values in increasing order and selecting the largest rank $k$ such that:
$$ p_k \leq \frac{k}{T} \alpha ,  $$
where $T$ is the total number of tests performed and $\alpha$ is the chosen significance level.
Then, all p-values until the $k$-th one included are deemed statistically significant.

In SVMIS, we apply the BH for each size separately.

\section{Sensitivity to the significance threshold $\alpha$}\label{app:sensitivity_alpha}

In the main text, we set the significance threshold of SVMIS to $\alpha=0.01$. 
To assess the extent to which the results of SVMIS are sensitive to the choice of $\alpha$, we perform a sensitivity analysis by repeating the benchmark procedure for different values of $\alpha$. 
Specifically, we considered values spanning four orders of magnitude, $\alpha \in [10^{-4},10^{-1}]$, while using the same configuration of parameters analyzed in the benchmark analysis.

\begin{figure}
    \centering
    \includegraphics[width=0.95\linewidth]{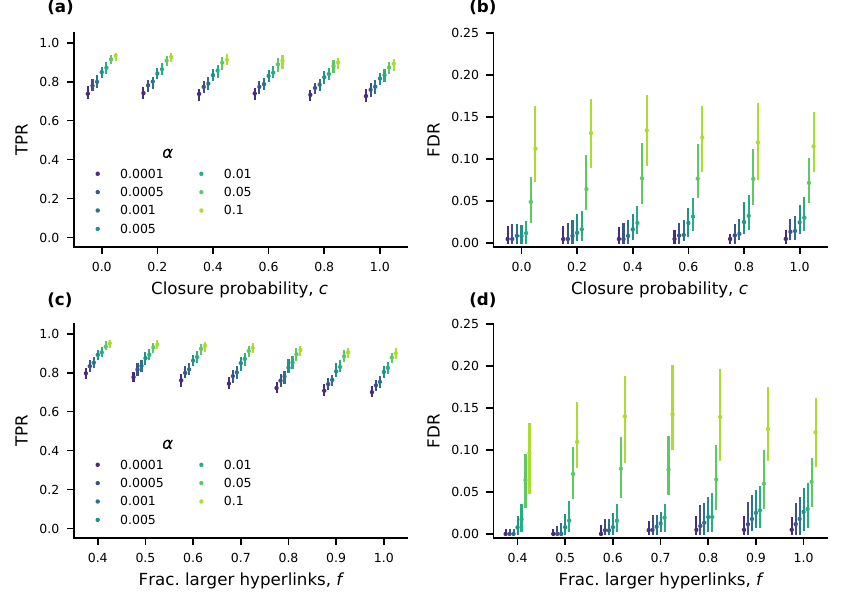}
    \caption{{\bf Sensitivity analysis to the value of $\alpha$.} 
    (a) True Positive Rate (TPR, y-axis) between the implanted sets and those detected by
    Statistically Validated Maximal Interacting Sets as a function of the closure probability $c$ with N=200 and $d=0.005$.
    Markers refer to the median across 100 realizations of the benchmark and error bars to the interval between the $10$-th and the $90$-th percentile. 
    Different values of $\alpha$ are shown with different colors.
    (b) False Discovery Rate (FDR, y-axis) for SVMIS as a function of $c$ for the same benchmark realizations. 
    (c) TPR (y-axis) for SVMIS as a function of the fraction of larger hyperedges $f$ over 100 benchmark realizations. 
    (d) FDR (y-axis) for SVMIS as a function of the fraction of larger hyperedges $f$ over the same benchmark realizations of (c).
    }
    \label{fig:alpha-sensitivity}
\end{figure}

We show the results in Fig.~\ref{fig:alpha-sensitivity} for the same configuration of parameters.
Increasing $\alpha$ leads to an increase in TPR, indicating that a larger fraction of the implanted sets is recovered. 
However, this gain in TPR comes with an increase in FDR, because a larger number of non-implanted sets are also validated. 
Conversely, decreasing $\alpha$ reduces FDR but also lowers TPR, as fewer implanted sets pass the statistical validation step.

Overall, the sensitivity analysis supports the use of $\alpha=0.01$ as a reasonable compromise between TPR and FDR. 
For values below $\alpha=0.01$, the method becomes increasingly conservative, with a progressive reduction in TPR. 
For values above $\alpha=0.01$, the increase in TPR comes with a marked increase in FDR, which rises from approximately $2\%$ to values above $5$--$10\%$, depending on the benchmark configuration. 
Larger values of $\alpha$ also lead to stronger fluctuations across benchmark realizations, as indicated by the wider error bars. 
Thus, $\alpha=0.01$ lies in a stable and interpretable region of the trade-off, maintaining high recovery of implanted sets while limiting the validation of spurious sets.

\bibliographystyle{unsrtnat}

\begin{thebibliography}{47}
	\providecommand{\natexlab}[1]{#1}
	\providecommand{\url}[1]{\texttt{#1}}
	\expandafter\ifx\csname urlstyle\endcsname\relax
	\providecommand{\doi}[1]{doi: #1}\else
	\providecommand{\doi}{doi: \begingroup \urlstyle{rm}\Url}\fi
	
	\bibitem[Newman(2003)]{newman2003structure}
	M.~E.~J. Newman.
	\newblock The structure and function of complex networks.
	\newblock \emph{SIAM Review}, 45:\penalty0 167--256, 2003.
	
	\bibitem[Boccaletti et~al.(Fervier 2006)Boccaletti, Latora, Moreno, Chavez, and
	Hwang]{boccaletti2006complex}
	S.~Boccaletti, V.~Latora, Y.~Moreno, M.~Chavez, and D-U. Hwang.
	\newblock Complex networks : {{Structure}} and dynamics.
	\newblock \emph{Phys. Rep.}, 424\penalty0 (4-5):\penalty0 175--308, Fervier
	2006.
	
	\bibitem[Battiston et~al.(2020)Battiston, Cencetti, Iacopini, Latora, Lucas,
	Patania, Young, and Petri]{battiston2020networks}
	Federico Battiston, Giulia Cencetti, Iacopo Iacopini, Vito Latora, Maxime
	Lucas, Alice Patania, Jean-Gabriel Young, and Giovanni Petri.
	\newblock Networks beyond pairwise interactions: structure and dynamics.
	\newblock \emph{Physics Reports}, 2020.
	
	\bibitem[Battiston et~al.(2021)Battiston, Amico, Alain, and
	et~al.]{battiston2021physics}
	Federico Battiston, Enrico Amico, Barrat Alain, and et~al.
	\newblock The physics of higher-order interactions in complex systems.
	\newblock \emph{Nature Physics}, 2021.
	
	\bibitem[Battiston et~al.(2025)Battiston, Capraro, Karimi, Lehmann, Migliano,
	Sadekar, S{\'a}nchez, and Perc]{battiston2025higher}
	Federico Battiston, Valerio Capraro, Fariba Karimi, Sune Lehmann,
	Andrea~Bamberg Migliano, Onkar Sadekar, Angel S{\'a}nchez, and Matja{\v{z}}
	Perc.
	\newblock Higher-order interactions shape collective human behaviour.
	\newblock \emph{Nature Human Behaviour}, pages 1--17, 2025.
	
	\bibitem[Patania et~al.(2017)Patania, Petri, and Vaccarino]{patania2017shape}
	Alice Patania, Giovanni Petri, and Francesco Vaccarino.
	\newblock The shape of collaborations.
	\newblock \emph{EPJ Data Sci.}, 6\penalty0 (1):\penalty0 18, 2017.
	
	\bibitem[Grilli et~al.(2017)Grilli, Barab{\'a}s, {Michalska-Smith}, and
	Allesina]{grilli2017higher}
	Jacopo Grilli, Gy{\"o}rgy Barab{\'a}s, Matthew~J {Michalska-Smith}, and Stefano
	Allesina.
	\newblock Higher-order interactions stabilize dynamics in competitive network
	models.
	\newblock \emph{Nature}, 548\penalty0 (7666):\penalty0 210, 2017.
	
	\bibitem[Jost and Mulas(2019)]{jost2019hypergraph}
	J{\"u}rgen Jost and Raffaella Mulas.
	\newblock Hypergraph {{Laplace}} operators for chemical reaction networks.
	\newblock \emph{Advances in Mathematics}, 351:\penalty0 870--896, 2019.
	
	\bibitem[Petri et~al.(2014)Petri, Expert, Turkheimer, {Carhart-Harris}, Nutt,
	Hellyer, and Vaccarino]{petri2014homological}
	Giovanni Petri, Paul Expert, Federico Turkheimer, Robin {Carhart-Harris}, David
	Nutt, Peter~J Hellyer, and Francesco Vaccarino.
	\newblock Homological scaffolds of brain functional networks.
	\newblock \emph{J. R. Soc. Interface}, 11\penalty0 (101):\penalty0 20140873,
	2014.
	
	\bibitem[Coscia and Neffke(2017)]{coscia2017network}
	Michele Coscia and Frank~MH Neffke.
	\newblock Network backboning with noisy data.
	\newblock In \emph{2017 IEEE 33rd International Conference on Data Engineering
		(ICDE)}, pages 425--436. IEEE, 2017.
	
	\bibitem[Grady et~al.(2012)Grady, Thiemann, and Brockmann]{grady2012robust}
	Daniel Grady, Christian Thiemann, and Dirk Brockmann.
	\newblock Robust classification of salient links in complex networks.
	\newblock \emph{Nature Communications}, 3\penalty0 (1):\penalty0 864, May 2012.
	\newblock ISSN 2041-1723.
	
	\bibitem[Mercier et~al.(2022)Mercier, Scarpino, and
	Moore]{mercier2022effective}
	Alexander Mercier, Samuel Scarpino, and Cristopher Moore.
	\newblock Effective resistance against pandemics: Mobility network
	sparsification for high-fidelity epidemic simulations.
	\newblock \emph{PLOS Computational Biology}, 18\penalty0 (11):\penalty0 1--17,
	11 2022.
	
	\bibitem[Imre et~al.(2020)Imre, Tao, Wang, Zhao, Feng, and
	Wang]{imre2020spectrum}
	Martin Imre, Jun Tao, Yongyu Wang, Zhiqiang Zhao, Zhuo Feng, and Chaoli Wang.
	\newblock Spectrum-preserving sparsification for visualization of big graphs.
	\newblock \emph{Computers \& Graphics}, 87:\penalty0 89--102, 2020.
	\newblock ISSN 0097-8493.
	
	\bibitem[Serrano et~al.(2009)Serrano, Bogun{\'a}, and
	Vespignani]{serrano2009extracting}
	M~{\'A}ngeles Serrano, Mari{\'a}n Bogun{\'a}, and Alessandro Vespignani.
	\newblock Extracting the multiscale backbone of complex weighted networks.
	\newblock \emph{Proceedings of the national academy of sciences}, 106\penalty0
	(16):\penalty0 6483--6488, 2009.
	
	\bibitem[Tumminello et~al.(2011)Tumminello, Micciche, Lillo, Piilo, and
	Mantegna]{tumminello2011statistically}
	Michele Tumminello, Salvatore Micciche, Fabrizio Lillo, Jyrki Piilo, and
	Rosario~N Mantegna.
	\newblock Statistically validated networks in bipartite complex systems.
	\newblock \emph{PloS One}, 6\penalty0 (3):\penalty0 e17994, 2011.
	
	\bibitem[Kirkley(2025)]{kirkley2025fast}
	Alec Kirkley.
	\newblock Fast nonparametric inference of network backbones for weighted graph
	sparsification.
	\newblock \emph{Phys. Rev. X}, 15:\penalty0 031013, Jul 2025.
	
	\bibitem[Tumminello et~al.(2012)Tumminello, Lillo, Piilo, and
	Mantegna]{tumminello2012identification}
	Michele Tumminello, Fabrizio Lillo, Jyrki Piilo, and Rosario~N Mantegna.
	\newblock Identification of clusters of investors from their real trading
	activity in a financial market.
	\newblock \emph{New Journal of Physics}, 14\penalty0 (1):\penalty0 013041,
	2012.
	
	\bibitem[Musciotto et~al.(2016)Musciotto, Marotta, Miccich{\`e}, Piilo, and
	Mantegna]{musciotto2016patterns}
	Federico Musciotto, Luca Marotta, Salvatore Miccich{\`e}, Jyrki Piilo, and
	Rosario~N Mantegna.
	\newblock Patterns of trading profiles at the nordic stock exchange. a
	correlation-based approach.
	\newblock \emph{Chaos, Solitons \& Fractals}, 88:\penalty0 267--278, 2016.
	
	\bibitem[Musciotto et~al.(2018{\natexlab{a}})Musciotto, Marotta, Piilo, and
	Mantegna]{musciotto2018long}
	Federico Musciotto, Luca Marotta, Jyrki Piilo, and Rosario~N Mantegna.
	\newblock Long-term ecology of investors in a financial market.
	\newblock \emph{Palgrave Communications}, 4\penalty0 (1):\penalty0 1--12,
	2018{\natexlab{a}}.
	
	\bibitem[Li et~al.(2014)Li, Palchykov, Jiang, Kaski, Kert{\'e}sz, Micciche,
	Tumminello, Zhou, and Mantegna]{li2014statistically}
	Ming-Xia Li, Vasyl Palchykov, Zhi-Qiang Jiang, Kimmo Kaski, J{\'a}nos
	Kert{\'e}sz, Salvatore Micciche, Michele Tumminello, Wei-Xing Zhou, and
	Rosario~N Mantegna.
	\newblock Statistically validated mobile communication networks: the evolution
	of motifs in european and chinese data.
	\newblock \emph{New Journal of Physics}, 16\penalty0 (8):\penalty0 083038,
	2014.
	
	\bibitem[Mantegna(1999)]{mantegna1999hierarchical}
	Rosario~N. Mantegna.
	\newblock Hierarchical structure in financial markets.
	\newblock \emph{The European Physical Journal B - Condensed Matter and Complex Systems}, 11:\penalty0 193--197, 1999.
	
	\bibitem[Tumminello et~al.(2005)Tumminello, Aste, Di~Matteo, and
	Mantegna]{tumminello2005tool}
	Michele Tumminello, Tomaso Aste, Tiziana Di~Matteo, and Rosario~N. Mantegna.
	\newblock A tool for filtering information in complex systems.
	\newblock \emph{Proceedings of the National Academy of Sciences}, 102\penalty0
	(30):\penalty0 10421--10426, 2005.
	
	\bibitem[Massara et~al.(2016)Massara, Di~Matteo, and Aste]{massara2016network}
	Guido~Previde Massara, Tiziana Di~Matteo, and Tomaso Aste.
	\newblock {Network Filtering for Big Data: Triangulated Maximally Filtered
		Graph}.
	\newblock \emph{Journal of Complex Networks}, 5\penalty0 (2):\penalty0
	161--178, 2016.
	
	\bibitem[Tumminello et~al.(2007)Tumminello, Coronnello, Lillo, Micciche, and
	Mantegna]{tumminello2007spanning}
	Michele Tumminello, Claudia Coronnello, Fabrizio Lillo, Salvatore Micciche, and
	Rosario~N. Mantegna.
	\newblock Spanning trees and bootstrap reliability estimation in
	correlation-based networks.
	\newblock \emph{International Journal of Bifurcation and Chaos}, 17\penalty0
	(7):\penalty0 2319--2329, 2007.
	
	\bibitem[Musciotto et~al.(2018{\natexlab{b}})Musciotto, Marotta, Miccich{\`e},
	and Mantegna]{musciotto2018bootstrap}
	Federico Musciotto, Luca Marotta, Salvatore Miccich{\`e}, and Rosario~N
	Mantegna.
	\newblock Bootstrap validation of links of a minimum spanning tree.
	\newblock \emph{Physica A: Statistical Mechanics and its Applications},
	512:\penalty0 1032--1043, 2018{\natexlab{b}}.
	
	\bibitem[Barrat et~al.(2004)Barrat, Barthélemy, Pastor-Satorras, and
	Vespignani]{barrat2004architecture}
	A.~Barrat, M.~Barthélemy, R.~Pastor-Satorras, and A.~Vespignani.
	\newblock The architecture of complex weighted networks.
	\newblock \emph{Proceedings of the National Academy of Sciences}, 101\penalty0
	(11):\penalty0 3747--3752, 2004.
	
	\bibitem[Musciotto et~al.(2021)Musciotto, Battiston, and
	Mantegna]{musciotto2021detecting}
	Federico Musciotto, Federico Battiston, and Rosario~N Mantegna.
	\newblock Detecting informative higher-order interactions in statistically
	validated hypergraphs.
	\newblock \emph{Communications Physics}, 4\penalty0 (218), 2021.
	
	\bibitem[Landry et~al.(2024{\natexlab{a}})Landry, Amburg, Shi, and
	Aksoy]{landry2024filtering}
	Nicholas~W Landry, Ilya Amburg, Mirah Shi, and Sinan~G Aksoy.
	\newblock Filtering higher-order datasets.
	\newblock \emph{Journal of Physics: Complexity}, 5\penalty0 (1):\penalty0
	015006, feb 2024{\natexlab{a}}.
	
	\bibitem[Kirkley et~al.(2025)Kirkley, Felippe, and
	Battiston]{kirkley2025structural}
	Alec Kirkley, Helcio Felippe, and Federico Battiston.
	\newblock Structural reducibility of hypergraphs.
	\newblock \emph{Physical Review Letters}, 135\penalty0 (24):\penalty0 247401, 2025.
	
	\bibitem[Lucas et~al.(2026)Lucas, Gallo, Ghavasieh, Battiston, and
	Domenico]{lucas2025reducibility}
	Maxime Lucas, Luca Gallo, Arsham Ghavasieh, Federico Battiston, and Manlio~De
	Domenico.
	\newblock Reducibility of higher-order networks from dynamics.
	\newblock \emph{Nature Communications}, 17\penalty0 (1):\penalty0 1551, 2026. 
		
	\bibitem[Ceria and Takes(2025)]{ceria2025relevance}
	Alberto Ceria and Frank~W. Takes.
	\newblock The relevance of higher-order ties.
	\newblock \emph{EPJ Data Science}, 14\penalty0 (1):\penalty0 62, Aug 2025.
	\newblock ISSN 2193-1127.
	
	\bibitem[Lotito et~al.(2022)Lotito, Musciotto, Montresor, and
	Battiston]{lotito2022higher}
	Q.~Francesco Lotito, Federico Musciotto, Alberto Montresor, and Federico
	Battiston.
	\newblock Higher-order motif analysis in hypergraphs.
	\newblock \emph{Communications Physics}, 5\penalty0 (79), 2022.
	
	\bibitem[LaRock and Lambiotte(2023)]{larock2023encapsulation}
	Timothy LaRock and Renaud Lambiotte.
	\newblock Encapsulation structure and dynamics in hypergraphs.
	\newblock \emph{Journal of Physics: Complexity}, 4\penalty0 (4):\penalty0
	045007, nov 2023.
	
	\bibitem[Landry et~al.(2024{\natexlab{b}})Landry, Young, and
	Eikmeier]{landry2024simpliciality}
	Nicholas~W. Landry, Jean-Gabriel Young, and Nicole Eikmeier.
	\newblock The simpliciality of higher-order networks.
	\newblock \emph{EPJ Data Science}, 13\penalty0 (1):\penalty0 17, Mar
	2024{\natexlab{b}}.
	\newblock ISSN 2193-1127.
	
	\bibitem[Lamata-Ot\'{\i}n et~al.(2025)Lamata-Ot\'{\i}n, Malizia, Latora,
	Frasca, and G\'omez-Garde\~nes]{lamata2025hyperedge}
	Santiago Lamata-Ot\'{\i}n, Federico Malizia, Vito Latora, Mattia Frasca, and
	Jes\'us G\'omez-Garde\~nes.
	\newblock Hyperedge overlap drives synchronizability of systems with
	higher-order interactions.
	\newblock \emph{Phys. Rev. E}, 111:\penalty0 034302, Mar 2025.
	
	\bibitem[Squartini et~al.(2015)Squartini, de~Mol, den Hollander, and
	Garlaschelli]{squartini2015breaking}
	Tiziano Squartini, Joey de~Mol, Frank den Hollander, and Diego Garlaschelli.
	\newblock Breaking of ensemble equivalence in networks.
	\newblock \emph{Physical Review Letters}, 115:\penalty0 268701, Dec 2015.
	
	\bibitem[Wang et~al.(2015)Wang, Zhao, and Zhang]{wang2015efficient}
	Minghui Wang, Yongzhong Zhao, and Bin Zhang.
	\newblock Efficient test and visualization of multi-set intersections.
	\newblock \emph{Scientific Reports}, 5\penalty0 (1):\penalty0 16923, Nov 2015.
	\newblock ISSN 2045-2322.
	
	\bibitem[Benjamini and Hochberg(1995)]{benjamini1995controlling}
	Yoav Benjamini and Yosef Hochberg.
	\newblock Controlling the false discovery rate: A practical and powerful
	approach to multiple testing.
	\newblock \emph{J R Statist Soc B}, 57\penalty0 (1):\penalty0 289--300, 1995.
	
	\bibitem[Kobayashi et~al.(2019)Kobayashi, Takaguchi, and
	Barrat]{kobayashi2019structured}
	Teruyoshi Kobayashi, Taro Takaguchi, and Alain Barrat.
	\newblock The structured backbone of temporal social ties.
	\newblock \emph{Nature Communications}, 10\penalty0 (1):\penalty0 1--11, 2019.
	
	\bibitem[Amburg et~al.(2020)Amburg, Veldt, and Benson]{amburg2020clustering}
	Ilya Amburg, Nate Veldt, and Austin Benson.
	\newblock Clustering in graphs and hypergraphs with categorical edge labels.
	\newblock In \emph{Proceedings of The Web Conference 2020}, WWW '20, page
	706–717, New York, NY, USA, 2020. Association for Computing Machinery.
	\newblock ISBN 9781450370233.
	
	\bibitem[Lotito et~al.(2026)Lotito, Betti, Nortier, Montresor, and
	Battiston]{lotito2026hypergraphxdata}
	Quintino~Francesco Lotito, Lorenzo Betti, Berné Nortier, Alberto Montresor,
	and Federico Battiston.
	\newblock Hypergraphx-data: a repository for higher-order network data.
	\newblock \emph{Journal of Complex Networks}, 2026.
	
	\bibitem[{FirstRate Data}()]{firstrate}
	{FirstRate Data}.
	\newblock Firstrate data — historical intraday market price data.
	\newblock \url{https://firstratedata.com/}.
	
	\bibitem[Gerig(2015)]{gerig2015high}
	Austin Gerig.
	\newblock High-frequency trading synchronizes prices in financial markets.
	\newblock \emph{Available at SSRN 2173247}, 2015.
	
	\bibitem[cul(Accessed Jan 2026)]{culinarydb}
	{CulinaryDB}.
	\newblock \url{https://cosylab.iiitd.edu.in/culinarydb/}, Accessed Jan 2026.
	
	\bibitem[Caprioli et~al.(2025)Caprioli, Kulkarni, Battiston, Iacopini, Santoro,
	and Latora]{caprioli2025networks}
	Claudio Caprioli, Saumitra Kulkarni, Federico Battiston, Iacopo Iacopini,
	Andrea Santoro, and Vito Latora.
	\newblock The networks of ingredient combinations as culinary fingerprints of
	world cuisines.
	\newblock \emph{npj Science of Food}, 9\penalty0 (1):\penalty0 242, Nov 2025.
	\newblock ISSN 2396-8370.
	
	\bibitem[Lotito et~al.(2023)Lotito, Contisciani, De~Bacco, Di~Gaetano, Gallo,
	Montresor, Musciotto, Ruggeri, and Battiston]{lotito2023hypergraphx}
	Quintino~Francesco Lotito, Martina Contisciani, Caterina De~Bacco, Leonardo
	Di~Gaetano, Luca Gallo, Alberto Montresor, Federico Musciotto, Nicol{\`o}
	Ruggeri, and Federico Battiston.
	\newblock Hypergraphx: a library for higher-order network analysis.
	\newblock \emph{Journal of Complex Networks}, 11\penalty0 (3):\penalty0
	cnad019, 2023.
	
\end{thebibliography}

\end{document}